\documentclass[10pt,oneside]{article}
\usepackage{amsfonts,amssymb,latexsym,amsthm,amsmath}
\usepackage[dvips]{graphics}

\newtheorem{theorem}{Theorem}[section]

\newtheorem{corollary}{Corollary}[section]

\newtheorem{definition}{Definition}[section]

\theoremstyle{remark}

\def \Z{\mathbb Z}

\def\C{\Bbb C}

\def \P{\mathcal P}
\def \C{\mathcal C}

\begin{document}

\title{Aspects of Nonabelian Group Based Cryptography:\\ A Survey and Open Problems}

\author{Benjamin Fine\\
Department of Mathematics\\
Fairfield University\\Fairfield, Connecticut 06430,
United States\\fine@mail.fairfield.edu \and
Maggie Habeeb\\
Department of Mathematics\\
City University of New York\\
New York, NY 10016,United States\\mhabeeb@gc.cuny.edu \and
Delaram Kahrobaei\\
Department of Computer Science\\
City University of New York\\New York,Ny 10016,
United States\\dkahrobaei@gc.cuny.edu \and
Gerhard Rosenberger\\
Fachbereich Mathematik\\
TU Dortmund\\
Dortmund 44227,
Germany\\gerhard.rosenberger@math.dortmund-uni.de}

\date{ }

\maketitle

\begin{abstract} Most common public key cryptosystems and public key exchange protocols presently in use, such as
the  RSA algorithm, Diffie-Hellman, and elliptic curve methods are number theory based and hence
depend on the structure of abelian groups.  The strength of computing machinery has made these
techniques theoretically susceptible to attack and hence recently there has been an active line of
research to develop cryptosystems and key exchange protocols using noncommutative cryptographic
platforms. This line of investigation has been given the broad title of
{\bf noncommutative algebraic cryptography}.  This was initiated by two public key protocols that used the braid groups, one by Ko, Lee et.al.and one by Anshel, Anshel and Goldfeld. The study of these protocols and the group theory surrounding them has had a large effect on research in infinite group theory.  In this paper we survey these noncommutative group based methods and discuss several ideas in abstract infinite group theory that have arisen from them. We then present a set of open problems.
\end{abstract}

\centerline {TABLE OF CONTENTS}

$\hphantom{xx}$ 1. Introduction and Nonabelian Group Cryptography\newline
$\hphantom{xx}$ 2. The Basics of Public Key Cryptography\newline
$\hphantom{xx}$ 3. Fundamentals of Free Group Cryptography\newline
$\hphantom{xx}$ 4. Public Key Exchange Using Nonabelian Groups\newline
$\hphantom{xx}$ 5. A Key Transport Protocol Based on a Shamir Three Pass\newline
$\hphantom{xx}$ 6. Digitial Signatures, Authentication and Password Verification\newline
$\hphantom{xx}$ 7. Braid Group Cryptography and General Platform Groups\newline
$\hphantom{xx}$ 8. Cryptography with Polycyclic Groups\newline
$\hphantom{xx}$ 9. Generic Complexity and Asymptotic Density\newline
$\hphantom{xx}$ 10. The Generic Free Group Property\newline
$\hphantom{xx}$ 11. Open Problems in Nonabelian Group Based Cryptography

\section{Introduction and Nonabelian Group Based Cryptography}

Traditionally {\bf cryptography} is the science and/or art of devising and implementing secret codes or {\bf cryptosystems}. {\bf Cryptanalysis} is the science and/or art of breaking cryptosystems while {\bf cryptology} refers to the whole field of cryptography plus cryptanalysis. In most modern literature cryptography is used synonomously with cryptology.  Presently    there is an increasing need for secure cryptosystems due to the use of internet shopping, electronic financial
transfers and so on.

Most common public key cryptosystems and public key exchange protocols presently in use, such as
the  RSA algorithm, Diffie-Hellman, and elliptic curve methods are number theory based and hence theoretically
depend on the structure of abelian groups. Although there have been no successful attacks on the standard protocols there is a feeling that the strength of computing machinery has made these techniques less secure. As a result of this there has been an active line of research to develop and analyze new cryptosystems and key exchange protocols based on noncommutative cryptographic platforms. This line of investigation has been given the broad title of {\bf noncommutative algebraic cryptography} (see [MSU]).

Up to this point the main sources for noncommutative cryptographic platforms has been nonabelian
groups.  In cryptosystems based on these objects algebraic properties of the platforms are used
prominently in both devising cryptosystems and in cryptanalysis. In particular the
difficulty, in a complexity sense, of certain algorithmic problems in finitely presented groups, such as the
conjugator search problem, has been crucial in encryption and decryption.

The main sources for nonabelian groups are combinatorial group theory and linear group theory.
Braid group cryptography (see [D]), where encryption is done within the classical braid groups, is one prominent example. The one way functions in braid group systems are based on the difficulty of solving group theoretic decision problems such as the
conjugacy problem and conjugator search problem.  Although braid group cryptography had initial spectacular success, various potential attacks have been identified.  Borovik, Myasnikov, Shpilrain [BMS] and others have studied the statistical aspects of these attacks and have identitifed what are termed black holes
in the platform groups outside of which present cryptographic problems.  Baumslag. Fine and Xu in [BFX] and [X] suggested potential cryptosystems using a combination of combinatorial group theory and linear groups and a general schema for these types of cryptosystems was given. In [BFX 2] a
public key version of this schema using the classical modular group as a platform was presented.
A cryptosystem using the the extended modular group $SL_2(\Bbb Z)$ was developed by Yamamura
([Y]) but was subsequently shown to have loopholes ([BG], [S], [HGS]).  In [BFX 2] attacks based on
these loopholes were closed.

The study and cryptanalysis of potential platform groups has had a strong positive effect on both group theory and complexity theory.  Motivated in large part by cryptography, there has been tremendous interest in {\bf asymptotic group theory} and {\bf generic properties} (see sections 9 and 10).

In this article we will present an overview of these combinatorial group theoretic methods. We will introduce free group cryptography and then the seminal Anshel-Anshel-Goldfeld and Ko-Lee Protocols.  We will further discuss how to use combinatorial group theory in digital signatures and password verification. Then we will discuss potential platform groups and give a brief review of braid group cryptography. Next we introduce asymptotic density and its application to cryptanalysis and evaluation of cryptosystems. In particular we will look at some theoretical results concerning the generic free group property. Finally a list of open problems (in no way exhaustive) in this area will be presented.

There are several recent books on this area. The most comprehensive is by Myasnikov, Shpilrain and Ushakov [MSU].  In this article we touch on several topics that are not mentioned in that book, for example polycylic group cryptography, digital signature protocols and password authentication and the generic free group property.

\section{Basics of Public Key Cryptography}

In this section we describe the standard terminology used in cryptography and then introduce the two most common public key methods, the Diffie-Hellman protocol and the RSA protocol.

In general both the {\bf plaintext message} (uncoded message) and the {\bf ciphertext message} (
coded message) are written in some $N$-letter alphabet which is usually the same for both plaintext
and code.  The method of coding or the encoding algorithm is then a transformation of the
$N$-letters.  The most common way to perform this transformation is to consider the $N$ letters as
$N$ integers modulo $N$ and then perform a number theoretical function on them.  Therefore most
encoding algorithms use modular arithmetic and hence cryptography is closely tied to number
theory.

Modern cryptography is usually separated into {\bf classical} or {\bf symmetric key cryptography} and {\bf public key cryptography}.  In the former, both the encoding and decoding algorithms are supposedly known only
to the sender and receiver, usually referred to as Bob and Alice. In the latter, the encryption
method is public knowledge but only the receiver knows how to decode.

 The process of putting the plaintext
message into code is called {\bf enciphering} or {\bf encryption} while the reverse process is
called {\bf deciphering} or {\bf decryption}. Encryption algorithms partition the plaintext and
ciphertext message into {\bf message units}.  These are single letters or pairs of letters or
more generally $k$-vectors of letters.  The transformations are done on these message units and
the encryption algorithm is a mapping from the set of plaintext message units to the set of
ciphertext message units.  Putting this into a mathematical formulation we let

$$ \P = \{\text {set of all plaintext message units } \}  \text { and }  $$
$$ \C = \{\text { set of all ciphertext message units }\} .$$
The encryption algorithm is then the application of an invertible function
$$ f:\P \mapsto \C.$$
The function $f$ is the {\bf encryption map}.  The inverse
$$f^{-1}:\C \mapsto \P$$ is the
{\bf decryption} or {\bf deciphering map}. The triple $\{\P,\C,f\}$, consisting of a set of
plaintext message units, a set of cipertext message units and an encryption map is called a {\bf
cryptosystem}.

Breaking a code is called {\bf cryptanalysis}.  An attempt to break a code is called an {\bf
attack}.  Most cryptanalysis starts with a statistical frequency analysis of the plaintext
language used.  Cryptanalysis depends also on a knowledge of the form of the
code, that is, the type of cryptosystem used (see [K] or [MSU]).

Most classical cryptosystems are number theoretically derived crytosystems. In applying a cryptosystem
to an $N$ letter alphabet we consider the letters as integers mod $N$. The encryption algorithms
then apply number theoretic functions and use modular arithmetic on these integers.

We usually do not use a single letter at a time but rather a sequence of $k$ letters. The $k$ letters are then a {\bf message unit}.  An encryption algorithm is then a function
$$f: \Bbb Z_N^k \mapsto \Bbb Z_N^m$$

Presently there are many instances where secure information must be sent over open communication
lines.  These include for example banking and financial transactions, purchasing items via credit
cards over the internet and similar things. This led to the development of {\bf public key
cryptography}. Roughly, in classical cryptography only the sender and receiver know the encoding
and decoding methods.  Further it is a feature of such cryptosystems that if the encrypting method is known the decrypting can be carried out.

 In {\bf public key cryptography} the encryption method is public knowledge but only the receiver knows how to decode.
More precisely in a classical cryptosystem once the encrypting algorithm is known the decryption
algorithm can be implemented in approximately the same order of magnitude of time.  In a public
the decryption algorithm is much more difficult to implement. This difficulty depends on the type of computing machinery used ( much as primality testing) and as computers get more powerful, new and more secure pulic key cryptosystems become
necessary.

The basic idea in a public key cryptosystem is to have a {\bf one-way function}.  That is a
function which is easy to implement but very hard to invert.  Hence it becomes simple to encrypt a
message but very hard, unless you know the inverse, to decrypt.

The standard model for a public key cryptosystem is the following.
Alice wants to send a message to Bob.  The encrypting map $f_A$ for
Alice is public knowledge as well as the encrypting map $f_B$ for Bob.  On the other hand the
decryption algorithms $f_A^{-1}$ and $f_B^{-1}$ are secret and known only to Alice and Bob
respectively.  Let $\cal P$ be the message Alice wants to send to Bob.  She sends
$f_Bf_A^{-1}(\cal P)$. To decode Bob applies first $f_B^{-1}$, which only he knows.  This gives him
$f_B^{-1}(f_Bf_A^{-1}({\cal P})) = f_A^{-1}({\cal P})$.  He then looks up $f_A$ which is publically
available and applies this $f_A(f_A^{-1}(\cal P)) = \cal P$ to obtain the message.

Alice sends $f_Bf_A^{-1}(\cal P)$ rather than  just $f_B(\cal P)$ for {\bf authentication},
that is being certain from Bob's point of view that the message really came from Alice.  Suppose
$\cal P$ is Alice's verification; signature, social security number etc..  If Bob receives
$f_B(\cal P)$ it could be sent by anyone since $f_B$ is public.  On the other hand since only
Alice supposedly knows $f_A^{-1}$ getting a reasonable message from $f_A(f_B^{-1}f_Bf_A^{-1}(\cal
P))$ would verify that it is from Alice. Applying $f_B^{-1}$ alone should result in nonsense.

Getting a reasonable one way function can be a formidable task. The most widely used (at
present) public key systems are based on the difficulty of inverting certain number theoretic functions. The first real public key protocol was developed in 1976 by Diffie and Hellman using the difficulty of the {\bf discrete log problem}.

In modular arithmetic it is easy to raise an element to a power but difficult to determine, given
an element, if it is a power of another element.
Specifically if $G$ is a finite group, such as
the cyclic multiplicative group of $\Z_p$ where $p$ is a prime, and $h = g^k$ for some $k$ then
the {\bf discrete log} of $h$ to the base $g$ is any integer $t$ with $h = g^t$.  The rough form
of the Diffie-Helman public key system is as follows.

Bob and Alice will use a classical
cryptosystem based on a key $k$ with $1 < k < q-1$ where $q$ is a prime.  It is the key $k$ that
Alice must send to Bob. Let $g$ be a multiplicative generator of $\Z_q^\star$.  Alice chooses an $a
\in \Z_q$ with $1 < a < q-1$. She makes public $g^a$.  Bob chooses a $b \in \Z_q^\star$ and makes
public $g^b$.  The secret key is $g^{ab}$. Both Bob and Alice, but presumably noone else, can
discover this key.  Alice knows her secret power $a$ and the value $g^b$ is public from Bob.  Hence
she can compute the key $g^{ab} = (g^b)^a$. The analogous situation holds for Bob.  An attacker
however only knows $g$, $g^a$ and $g^b$. Unless the attacker can solve the discrete log problem, that
is finding $a$ or $b$ the key exchange is secure.

Notice that this depends upon
$$(g^a)^b = (g^b)^a$$
As we will see, the Ko-Lee protocol exactly mimics the Diffie-Hellman protocol within a nonabelian group by interpreting powers as conjugation.

In 1997 it became known that the ideas of public key cryptography were developed by British Intelligience Services prior to Diffie and Hellman.

In 1977 Rivest, Adelman and Shamir developed the {\bf RSA Algorithm} which is presently
the most widely used public key cryptosystems. It is based on the difficulty of factoring large
integers and in particular on the fact that it is easier to test for primality  than to factor. In basic outline at the simplest level it works as follows.

Alice chooses two large primes
$p_A,q_A$ and an integer $e_A$ relatively prime to $\Phi(p_Aq_A) = (p_A-1)(q_A-1)$.  It is
assumed that these integers are chosen randomly to minimize attack.
  The primes
she chooses should be quite large.  Originally RSA used primes of approximately 100 decimal
digits, but as computing and attack have become more sophisticated, larger primes have had to be
utilized.  Once Alice has obtained $p_A,q_A,e_A$ she lets $n_A =
p_Aq_A$ and computes $d_A$ , the multiplicative inverse of $e_A$ modulo $\Phi(n_A)$. That is
$d_A$ satisfies $e_Ad_A \equiv 1$ mod $(p_A-1)(q_A-1)$. She makes public the enciphering key
$K_A = (n_A,e_A)$ and the encryption algorithm known to all is
$$f_A(\P) = \P^{e_A} \text { mod } n_A$$
where $\P \in \Z_{n_A}$ is a message unit.

 It can be shown that if $(e_A,(p_A-1)(q_A-1)) = 1$
and $e_Ad_A \equiv 1$ mod $(p_A-1)(q_A-1)$ then $\P^{e_Ad_A} \equiv \P$ mod $n_A$
  Therefore the decryption algorithm is
$$f_A^{-1}(\C) = \C^{d_A} \text { mod } n_A.$$
Notice then that $f_A^{-1}(f_A(\P)) = \P^{e_Ad_A} \equiv \P$ mod $n_A$ so it is the inverse.

Now Bob makes the same type of choices to obtain $p_B,q_B,e_B$. He lets $n_B = p_Bq_B$ and makes
public his key $K_B = (n_B,e_B)$.

If Alice wants to send a message to Bob that can be authenticated to be from Alice she sends
$f_B(f_A^{-1}(\P))$.  An attack then requires factoring $n_A$ or $n_B$ which is much more difficult
than obtaining the primes $p_A,q_A,p_B,q_B$.

Again notice the use of commutativity.

There have been many extensions and enhancements of these basic public key protocols.  Elliptic curve cryptography uses the discrete log problem within the group of an elliptic curve.  This group is a finite abelian group and has certain advantages over the cyclic groups used in the standard Diffie-Hellman protocol. The book by Koblitz [K] has a thorough description of elliptic curve methods.

The El-Gamal cryptosystem is a technique to use the Diffie-Hellman key exchange method to do encryption. The method works as follows.  Suppose that Bob and Alice want to communicate openly.  They have exchanged a secret key $k$ that supposedly only they know.  Let $f_k$ be an encryption function or encryption algorithm based on the key $k$.  Alice wants to send the message $m$ to Bob and $m$ is given as a binary bit string. Alice sends to Bob
$$ f_k(m) \oplus k$$
where $k$ is a bit string for the key $k$ and $\oplus$ is addition modulo 2.

Bob knows the key $k$ and hence can compute it as a bibary string. He now computes
$$ f_k(m) \oplus k \oplus k$$
Since addition modulo 2 has order 2 we have
$$  f_k(m) \oplus k \oplus k = f_k(m).$$
Bob now applies the decryption algorithm $f_K^{-1}$ to decode the message. In practice a {\bf hash function} is usually applied to $k$ (see [B]).

\section{The Fundamentals of Free Group Cryptography}

The extension of all these ideas to noncommutative platforms is the subject of {\bf noncommutative algebraic cryptography}.  This involves the following ideas,
\smallskip

$\hphantom{xx}$ (1) General Algebraic Techniques for Developing Cryptosystems

$\hphantom{xx}$ (2) Potential Algebraic Platforms (Specific Groups, Rings, Etc.) for implementing the Techniques

$\hphantom{xx}$ (3) Cryptanalysis and Security Analysis of the Resulting Systems
\smallskip

The main source for noncommutative platforms are nonabelian groups and the main method for handling nonabelian groups in cryptography is combinatorial group theory.  This refers to the branch of group theory that studies groups by using group presentations, that is sets of generators and relations between them.
The basic idea in using combinatorial group theory for cryptography is that
elements of groups can be expressed as words in some alphabet. If there is an easy method to rewrite
group elements in terms of these words and further the technique used in this rewriting process can
be  supplied by a secret key then a cryptosystem can be created.  The simplest example is perhaps a
{\bf free group cryptosystem}.  This can be described in the following manner.

Consider a free group $F$ on free generators $x_1,...,x_r$. Then each element $g$ in $F$ has a unique
expression as a word $W(x_1,...,x_r)$.  Let $W_1,...,W_k$  with $W_i = W_i(x_1,...,x_r)$ be a set of
words in the generators $x_1,...,x_r$ of the free group $F$.  At the most basic level, to construct
a cryptosystem, suppose that we have a plaintext alphabet $\cal A$. For example suppose that ${\cal A} =
\{a,b,...\}$ are the symbols needed to construct meaningful messages in English. To encrypt,
use a substitution ciphertext  $${\cal A} \mapsto \{ W_1,...,W_k\}.$$
That is
$$ a \mapsto W_1, b \mapsto W_2,....$$
 Then given an  word
$W(a,b,...)$ in the plaintext alphabet form the free group word $W(W_1,W_2,....)$.  This represents
an element $g$ in $F$.  Send out $g$ as the secret message.

 In order to implement this scheme we need a concrete representation of $g$ and
then for decryption a way to rewrite $g$ back in terms of $W_1,...,W_k$.
This concrete representation is the idea behind {\bf homomorphic cryptosystems}.

 The decryption algorithm in a free group cryptosystem then depends on the {\bf Reidemeister-Schreier rewriting process}.  This is a method to rewrite elements of a subgroup of a free group in terms of the generators of that subgroup.  We refer to [MKS] or [GB 1] for a complete description of the technique. Roughly it works as follows. Assume that $W_1,...,W_k$  are free generators for some subgroup
$H$ of a free group $F$ on $\{x_1,...,x_n\}$. Each $W_i$ is then a reduced word in the generators $\{x_1,...,x_n\}$.  A {\bf Schreier transversal} for $H$ is a set $\{h_1,...,h_t,...\}$ of (left) coset
representatives for $H$ in $F$ of a special form (see [MKS]).  Any
subgroup of a free group has a Schreier transversal. The Reidemeister-Schreier process allows one to
construct a set of generators $W_1,...,W_k$ for $H$ by using a Schreier transversal.  Further given
the Schreier transversal from which the set of generators for $H$ was constructed, the{ \bf
Reidemeister-Schreier Rewriting Process} allows us to algorithmically rewrite an element of $H$.
Given such an element expressed as a word $W = W(x_1,...,x_r)$ in the generators of $F$ this
algorithm rewrites $W$ as a word $W^\star(W_1,...,W_k)$ in the generators of $H$.

The knowledge of a Schreier transversal and the use of Reidemeister-Schreier
rewriting facilitates the decoding process in the free group case but is not essential.
Given a known set of generators for a subgroup the Stallings Folding Method to develop a
subgroup graph can also be utilized to rewrite in terms of the given generators.  The paper
by Kapovich and Myasnikov [KM] is now a standard reference for this method in free groups.
At present there is an ongoing study of the complexity of Reidemeister-Schreier being done by Baumslag, Brukhov, Fine and Troeger.

Pure free group cryptosystems are subject to various attacks and can be broken easily.  However a
public key free group cryptosystem using a free group representation in the Modular group was
developed by Baumslag, Fine and Xu [BFX 1 2]. The most successful attacks on free group cyrpotsystems are called {\bf length based attacks}.  Here an attacker multiplies a word in ciphertext by a generator to get a shorter word which could possibly be decoded.

Baumslag, Fine and Xu in [BFX 1] described the following general encryption scheme using free group cryptography. A further enhancement was discussed in the paper [BFX 2].

We start with  a finitely presented group
$$G = <X|R>$$ where $X =\{x_1,...,x_n\}$ and a faithful representation
$$\rho:G \mapsto \overline{G}.$$
$\overline{G}$ can be any one of several different kinds of objects - linear group, permutation group, power
series ring etc.

We assume that there is an algorithm to re-express an element of $\rho(G)$ in $\overline{G}$ in
terms of the generators of $G$. That is if $g = W(x_1,...,x_n,...) \in G$ where $W$ is a word in
the these generators and we are given $\rho(g) \in \overline{G}$ we can algorithmically find $g$ and
its expression as the word $W(x_1,...,x_n)$.

Once we have $G$ we assume that we have two free subgroups $K,H$ with $$H \subset K \subset G.$$
We assume that we have fixed Schreier transversals for $K$ in $G$ and for $H$ in $K$ both of
which are held in secret by the communicating parties Bob and Alice. Now based on the fixed Schreier transversals we have sets of Schreier
generators constructed from the Reidemeister-Schreier process
for $K$ and for $H$. $$ k_1,...k_m,... \hphantom{xx} \text { for } K$$
and
$$ h_1,...,h_t,...\hphantom{xx}\text { for } H.$$

Notice that the generators for $K$ will be given as words in $x_1,...,x_n$, the generators of $G$
while the generators for $H$ will be given as words in the generators $k_1,k_2,....$ for $K$. We note
further that $H$ and $K$ may coincide and  that $H$ and $K$ need not in general be free but only have
a unique set of normal forms so that the representation of an element in terms of the given Schreier
generators is unique.

We will encode within $H$, or more precisely within $\rho(H)$. We assume that
the number of generators for $H$ is larger than the set of characters within our plaintext
alphabet. Let ${\cal A } = \{a,b,c...\}$ be our plaintext alphabet. At the simplest level we choose
a starting point $i$, within the generators of $H$, and enclode
$$ a \mapsto h_i, b \mapsto h_{i+1},.... \text { etc. }$$

Suppose that Bob wants to communicate the message $W(a,b,c...)$ to Alice where $W$ is a word in
the plaintext alphabet. Recall that both Bob and Alice know the various Schreier transversals which
are kept secret between them. Bob then encodes $W(h_i,h_{i+1}...)$ and computes in $\overline{G}$ the
element $W(\rho(h_i),\rho(h_{i+1}),..)$ which he sends to Alice. This is sent as a matrix if
$\overline{G}$ is a linear group or as a permutation if $\overline{G}$ is a permutation group and so on.

Alice uses the algorithm for $\overline{G}$ relative to $G$ to rewrite $W(\rho(h_i),\rho(h_{i+1}),..)$
as a word $W^\star(x_1,...x_n)$  in the generators of $G$.  She then uses the Schreier
transversal for $K$ in $G$ to rewrite, using the Reidemeister-Schreier process, $W^\star$ as a
word $W^{\star\star}(k_1,...,k_s..)$  in the generators of $K$.  Since $K$ is free or
has  unique normal forms this expression for the element of $K$ is unique.  Once she has the
word written in the generators of $K$ she uses the transversal for $H$ in $K$ to rewrite again,
using the Reidemeister-Schreier process, in terms of the generators for $H$.  She then has a word
$W^{\star\star\star}(h_i,h_{i+1},...)$ and using $h_i \mapsto a, h_{i+1} \mapsto b,...$
decodes the message.

In actual implementation an additional {\it random noise factor} is added.

In [FBX 1,2] an inplementation of this process was presented that used for the base
group $G$ the classical modular group $M = PSL(2,\Bbb Z)$.  Further it was a polyalphabetic
cipher which was secure.

The system in the modular group $M$ was presented as follows. A list of finitely generated free subgroups $H_1,...,H_m$ of $M$ is public and presented by their systems of generators (presented as matrices).  In a full
practical implementation it is assumed that $m$ is large.  For each $H_i$ we have a Schreier
transversal  $$h_{1,i},...,h_{t(i),i}$$
and a corresponding ordered set of generators
$$W_{1,i},...,W_{m(i),i}$$
constructed from the Schreier transversal by the Reidemeister-Schreier process. It is
assumed that each $m(i) >> l$ where $l$ is the size of the plaintext alphabet, that is each
subgroup has many more generators than the size of the plaintext alphabet. Although Bob and
Alice know these subgroups in terms of free group generators, what is made public are generating
systems given in terms of matrices.

The subgroups on this list and their corresponding Schreier transversals can be chosen in a
variety of ways.  For example the commutator subgroup of the Modular group is free of rank 2
and some of the subgroups $H_i$ can be determined from homomorphisms of this subgroup onto a set
of finite groups.

Suppose that Bob wants to send a message to Alice.  Bob first chooses  three integers $(m,q,t)$
where
$$ m = \text { choice of the subgroup } H_m$$
$$ q = \text { starting point among the generators of } H_m $$$$\text { for the substitution of the
plaintext alphabet}$$
$$ t = \text { size of the message unit }.$$
We clarify the meanings of $q$ and $t$. Once Bob chooses $m$, to further clarify the meaning of
$q$, he makes the substitution  $$ a \mapsto W_{m,q}, b \mapsto W_{m,{q+1}},.....$$
Again the assumption is that $m(i) >> l$ so that starting almost anywhere in the sequence of
generators of $H_m$ will allow this substitution.  The message unit size $t$ is the number of coded
letters that Bob will place into each coded integral matrix.

Once Bob has made the choices $(m,q,t)$ he takes his plaintext message $W(a,b,...)$ and groups
blocks of $t$ letters. He then makes the given substitution above to form the
corresponding matrices in the Modular group;  $$ T_1,...,T_s.$$
We now introduce a {\it random noise factor}. After forming $T_1,...,T_s$ Bob then multiplies on the
right each $T_i$ by a random matrix in $M$, say $R_{T_i}$ ( different for each $T_i$).
The only restriction on this random matrix $R_{T_i}$ is that there is no free
cancellation in forming the product $T_iR_{T_i}$.  This can be easily checked and ensures that
the freely reduced form for $T_iR_{T_i}$  is just the concatenation of the expressions for $T_i$
and $R_{T_i}$.  Next he sends Alice the integral key $(m,q,t)$ by some public key method (RSA,
Anshel-Goldfeld etc.). He then sends the message as $s$ random matrices $$
T_1R_{T_1},T_2R_{T_2},...,T_sR_{T_s}.$$ Hence what is actually being sent out are not elements of
the chosen subgroup $H_m$ but rather elements of random right cosets of $H_m$ in $M$. The
purpose of sending coset elements is two-fold. The first is to hinder any geometric attack by
masking the subgroup.  The second is that it makes the resulting words in the Modular Group
generators longer, effectively hindering a brute force attack.

To decode the message Alice first uses public key decryption to obtain the integral keys $(m,q,t)$.
She then knows the subgroup $H_m$, the ciphertext substitution from the generators of $H_m$ and how
many letters $t$ each matrix encodes.  She next uses the algorithms described above to
express each $T_iR_{T_i}$ in terms of the generators of $M$.  She has knowledge of the Schreier transversal, which is held secretly by
Bob and Alice, so now uses the Reidemeister-Schreier rewriting process to start expressing
this freely reduced word in terms of the generators of $H_m$.  The Reidemeister-Schreier
rewriting is done letter by letter from left to right (see [MKS]).  Hence when she reaches $t$ of the free
generators she stops. Notice that the string that she is rewriting is longer than what she needs
to rewrite in order to decode  as a result of the random matrix $R_{T_i}$.  This is due to the
fact that she is actually rewriting not an element of the subgroup but an element in a right
coset.  This presents a further difficulty to an attacker. Since these are random right cosets it
makes it difficult to pick up statistical patterns in the generators even if more than one
message is intercepted.  In practice the subgroups should be changed with each message.

The initial key $(m,q,t)$ is changed frequently. Hence as mentioned above this method becomes
a type of polyalphabetic cipher. Polyalphabetic ciphers have historically been very difficult
to decode.

 A further variation of this method using the Magnus representation in a formal power series ring in noncommuting variables over a field was described in [BBFR].

\section{Public Key Exchange Using Nonabelian Groups}

Among the first attempts to use nonabelian groups in cryptography were the schemes of Anshel-Anshel-Goldfeld[AAG] and Ko-Lee et.al.[KoL].  Both sets of authors, at about the same time, proposed using nonabelian groups and combinatorial group theory for public key exchange. The security of these systems depended on the difficulty of solving certain "hard" group theoretic problems.

The methods of both Anshel-Anshel-Goldfeld and Ko-Lee can be considered as group theoretic analogs of the number theory based Diffie-Hellman method. The basic underlying idea is the following. If $G$ is a group and $g,h \in G$ we let $g^h$ denote the conjugate of $g$ by $h$, that is $g^h = h^{=1}gh$. The simple observation is that this behaves like ordinary exponentiation in that $(g^{h_1})^{h_2} = g^{h_1h_2}$.  From this straightforward idea one can exactly mimic the Diffie-Hellman protocol within a nonabelian group.

Both the Anshel-Anshel-Goldfeld protocol and the Ko-Lee protocol start with a platform group $G$ given by a group presentation. A major assumption in both protocols is that the elements of $G$ have nice unique normal forms that are easy to compute for given group elements.  However it is further assumed that given normal forms for $x,y \in G$, the normal form for the product $xy$, does not reveal $x$ or $y$.

We describe the Anshel-Anshel-Goldfeld public key exchange protocol first. Let $G$ be the platform group given by a finite prsentation and with the assumptions on normal forms as described above.

Alice and Bob want to communicate a shared secret. First,  Alice and Bob choose  random finitely generated subgroups of $G$ by giving a set of generators for each.
$$A = \{a_1,...,a_n\}, B= \{b_1,...,b_m\}$$
and make them public. The subgroup $A$ is Alice's subgroup while the subgroup $B$ is Bob's subgroup.

Alice chooses a secret group word $a = W(a_1,...,a_n)$ in her subgroup while Bob chooses a secret group word $b = V(b_1,...,b_m)$ in his subgroup. For an element $g \in G$ we let $NF(g)$ denote the normal form for $g$.    Alice knows her secret word $a$ and knows the generators $b_i$ of Bob's subgroup. She makes public the normal forms of the conjugates
$$ NF(b_i^{a}),i = 1,...,m.$$
Bob knows his secret word $b$ and the geerators $a_i$ of Alice's subgroup and makes public the normal forms of the conjugates
$$NF(a_j^b), j = 1,...,n).$$
The common shared secret is the commutator
$$[a,b] = a^{-1}b^{-1}ab = a^{-1}a^b = (b^a)^{-1}b$$

Notice that Alice knows $a^b$ since she knows $a$ in terms of generators $a_i$ of her subgroup and she knows the conjugates by $b$ since Bob has made the conjugates of the generators of $A$ by $b$ public. Since Alice knows $a^b$ she knows $[a,b] = a^{-1}a^b$.

In an analogous manner Bob knows $[a,b] = (b^a)^{-1}b$. An attacker would have to know the corresponding {\bf conjugator}, that is the element that conjugates each of the generators.  Given elements $g,h$ in a group $G$ where it is known that $g^k = k^{-1}gk = h$ the {\bf conjugator search problem} is to determine the conjugator $k$.  It is known that this problem is undecidable in general, that is there are groups where the conjugator cannot be determined algorithmically.  On the other hand there are groups where the conjugator search problem is solvable but "difficult", that is the complexity of solving the conjugator search problem is hard.  Such groups become the ideal platform groups for the Anshel-Anshel-Goldfeld protocol.

The security in this system is then in the difficulty of the {\bf conjugator search problem}.
Anshel, Anshel, Goldfeld suggested the Braid Groups as potential platforms and use for example $B_{80}$ with 12 or more generators in the subgroups. Their suggestion and that of Ko and Lee led to development of {\bf braid group cryptography}. There have been various attacks on the Braid group system. However some have been handled by changing the parameters.  In general the ideas remain valid despite the attacks. We will discuss this further in section 7.

Ko, Lee et. al. [KoL] developed a similar system that is a direct translation of the Diffie-Hellman protocol to a nonabelian group theoretic setting. Its security is based on the difficulty of the {\bf conjugacy problem}. We again assume that the platform group has nice unique normal forms that are easy to compute for a given group element but hard to recover the group element. Recall again that $g^h$ means the conjugate of $g$ by $h$

In the Ko-Lee protocol, Alice and Bob choose commuting subgroups $A$ and $B$ of the platform group $G$.  $A$ is Alice's subgroup while Bob's subgroup is $B$ and these are secret.  Now they completely mimic the classical Diffie-Hellman technique. There is a public element $g\in G$,  Alice chooses a random secret element $a\in A$ and makes public
$g^a$.  Bob chooses a random secret element $b \in B$ and makes public $g^b$.
 The secret shared key is $g^{ab}$. Notice that $ab = ba$ since the subgroups commute. It follows then that $(g^a)^b = g^{ab} = g^{ba} = (g^b)^a$ just as if these were exponents.  Hence both Bob and Alice can determine the common secret. The difficulty is in the difficulty of the conjugacy problem.

The {\bf conjugacy problem} for a group $G$, or more precisely for a group presentation for $G$, is given $g,h \in G$ to determine algorithmically if they are conjugates.  As with the conjugator search problem it is known that the conjugacy is undecidable in general but there are groups where it is but hard.  These groups then become the target platform groups for the Ko-Lee protocol. As with the Anshel-Anshel-Goldfeld protocol, Ko and Lee suggest the use of the Braid groups.

The conjugacy problem and the conjugator search problem are only two of the group theoretic search and decision problems that have been employed to construct one way functions in a cryptogrpahic setting.  We recall several other important such problems and then reference their use in encryption and public key exchange .

\begin{definition} [Word Problem]
Given a finitely presented group $G$, does there exist an algorithm
to decide whether or not a word in the generators is the trivial
word?
\end{definition}
\begin{definition}[Decision Conjugacy Problem]
Given a group $G$ with a finite presentation, does there exist an
algorithm to decide whether or not an arbitrary pair of words $u$
and $v$ in the generators of $G$ are conjugate? That is, is there an
$x\in G$ such that $x^{-1}ux=v$?
\end{definition}

\begin{definition}[Decomposition Problem]
Let $G$ be a finitely presented group with subgroups $A,B \leq G$.
Given two elements $u$ and $v$ of $G$, is there an algorithm to find
two elements $a\in A$ and $b\in B$ such that $aub=v$?
\end{definition}

\begin{definition}[Simultaneous Search Conjugacy problem]
Let $G$ be a finitely presented group. Given $u_{1},\cdots, u_{k},
v_{1},\cdots, v_{k}\in G$ with $x^{-1}u_{i}x=v_{i}$ for each
$i\in\left\{1,2, \cdots, k\right\}$, is there an algorithm to find
$z\in G$ satisfying $z^{-1}u_{i}z=v_{i}$ for each
$i\in\left\{1,2,...,k\right\}$?
\end{definition}

We now reference some uses for these problems. Wagner, Birget, Magliaveras and Sramka developed key exchange protocols based on the word problem. Kurt developed a protocol based on the decomposition problem.  Shpilrain and Ushakov developed a protocol based on the twisted conjugacy problem while Shpilrain and Zapata developed several encryption protocols based on various decision problems. Complete descriptions of these protocols can be found in [MSU].  More recently Anshel and Kahrobaei developed a noncommutative analog of the Cramer-Shoup key exchange method [AK].

We close this section by describing a noncommutative analog of the El Gamal public key exchange system based on the search conjugacy problem.  It was proposed by Kahrobaei and Khan [KKh]. As with the Ko-Lee and Anshel-Anshel-Goldfeld protocols we start with a finitely presented platform group $G$ given by a group presentation. As before the major assumptions  are that  the elements of $G$ have nice unique normal forms that are easy to compute for given group elements.  However it is further assumed that given normal forms for $x,y \in G$ the normal form for the product $xy$ does not reveal $x$ or $y$. Further $G$ contains two commuting finitely generated proper subgroups $S$ and $T$.
The cryptographic goal is for  Alice and Bob to establish a session key over an
unsecured network.

Bob chooses a secret element $s\in S$ and an arbitrary element $b\in
G$.  Bob publishes $b$ and $c=b^{s}$.  Suppose Alice wants to send
$x\in G$ as a session key to Bob. Then,
\begin{enumerate}
\item Alice chooses a random $t\in T$ and sends $E=x^{(c^{t})}$ to Bob along with the header $h=b^{t}$.
\item Bob then calculates $(b^{t})^{s}=(b^{s})^{t}=c^{t}$.
\item Now, Bob may calculate $E'=(c^{t})^{-1}$, allowing him to decrypt the session key since $(x^{(c^{t})})^{E'}=(x^{(c^{t})})^{(c^{t})^{-1}}=x$.
\end{enumerate}
The feasibility of this scheme relies on the assumption that
products and inverses in $G$ can be computed efficiently.
Determining Bob's private key $s$ entails solving the search
conjugacy problem for $G$. That is given $c$, $b$, and $c=b^{s}$,
determine $s$. Hence, the security of this scheme is based on the
assumption that there is no practical algorithm for solving the
search conjugacy problem for $G$.

A second El Gamal analog based on the search power conjugay problem was also proposed by Kahrobaei and Khan (see [KKh]).

\section {The Shamir Three Pass and Key Transport Protocols}

A {\bf key transport protocol} is a method that allows the sending of a key (telling for example what encryption system to use) from one user to another over a public airway. A group theoretic key transport protocol based on the Diffie-Hellman scheme can be developed in the following manner.
Suppose that we have a finitely presented group $G$ with the same assumptions made as in the Anshel-Anhsel-Goldfeld and Ko-Lee protocols.  That is, $G$ is given by a presentation and the elements of $G$ have nice normal forms.  Further it is assumed
 that $G$ has two large subgroups $A_1$,$A_2$ that commute elementwise.  Alternatively we
could use one large abelian subgroup $A$ of $G$. The meaning of large is of course hazy but here
means that within $G$ it is difficult to determine when an arbitrary element is in $A$ (or
$A_1,A_2$) and further $A$ (or $A_1,A_2$) is large enough so that random choices can be made from
them.

Now suppose that Bob wants to communicate with Alice via an open airway. The secret key telling
them which encryption system to use is encoded within the finitely generated group $G$ with the
properties given above.  The two subgroups $A_1,A_2$ which commute elementwise are kept secret by
Bob and Alice. $A_1$ is the subgroup for Bob and $A_2$ the subgroup for Alice. Bob wants to send
the key $W \in G$ to Alice.  He chooses two random elements $B_1,B_2 \in A_1$ and sends Alice the
message ( in encrypted form) $B_1WB_2$.  Alice now chooses two random elements  $C_1,C_2 \in
A_2$  and sends $C_1B_1WB_2C_2$ back to Bob. These messages appear in the representation of $G$
and hence for example as matrices or as reduced words in the generators so they don't appear as
solely concatenation of letters.  Since $A_1$ commutes elementwise with $A_2$ we have
$$C_1B_1WB_2C_2 = B_1C_1WC_2B_2.$$  Further since Bob knows his chosen elements $B_1$ and $B_2$
he can multiply by their inverses to obtain $C_1WC_2$ which he then sends back to Alice.  Since
Alice knows her chosen elements $C_1,C_2$ she can multiply by their inverses to obtain the key
$W$.  It is assumed that  for each message Bob and Alice would choose different pairs of random
elements from either $A_1$ or $A_2$. This method is known as a {\bf Shamir Three-Pass} which was  introduced by Shamir for general algebraic objects.

Notice that although this is roughly based on the Diffie-Hellman method it is not symmetric in
the communicating parties.  In the present scheme the secret key is completely determined by Bob,
who then communicates it to Alice.  The scheme then falls into the class of key transport
protocols rather than public key exchange protocols.  Key transport protocols are in most cases
designed assuming that an underlying encryption system (and usually also a signature verification
system) is in place.  The security of the key transport protocol will rely on the security of
these auxillary schemes.  In the group theoretic proposal the encryption
scheme is suggested to be done within the same group as the key transport protocol alhtough this
is not essential.  In the group theortic key transport protocol an attacker has knowledge of
the overall group $G$ and a view of encrypted messages.  The security lies in the difficulty of
determining the elementwise commuting subgroups $A_1,A_2$, which are kept secret by Bob and
Alice, and in the security of the actual encryption scheme.

A group $G$ is a candidate platform group for this type of key transport protocol if it has either a nice finite presentation $G = <X,R>$ with workable normal forms and has either a large abelian
subgroup $A$ or two large subgroups $A_1$,$A_2$ that commute elementwise. Although the word large
here is ambiguous we mean large enough so that random choices can be made from them. In
particular, for example, cyclic subgroups are inappropriate.  There also should be some tie
between the group used for the key exchange and the encryption method although this is not
essential. The standard braid groups, that will be described in the section 7 have several possibilitiess for normal forms and have large commuting subgroups.  Hence they are excellent candidates for this method.  In [BCFRX] several additional potential platform groups were suggested. These include the full automorphism group of a finitely generated free group, the matrix group $SL(4,\Bbb Z)$ and the surface braid groups.  Shpilrain and Ushakov [SU] used this method employing Thompson's group $F$ as a platform.  A length based attack on their system was attempted by Tsaban (see [MSU]). Further work on this method in the surface braid groups was done by Camps[C].

\section{Digital Signatures, Authentication and Password Security}

{\bf Authentication} is the process of determining that a message, supposedly from a given person, both does come from that person and has not been tampered with. Authentication plays a major role in transmitting encrypted messages. Often this takes the form of a {\bf digital signature}. A
signature scheme provides a way for each user to sign messages so
that the signatures can later be verified by anyone else. More
specifically, each user can create a matched pair of private and
public signature for the message (using the signer's public key).
The verifier can convince himself that the message contents have not
been altered since the message was signed. Also the signer cannot
later repudiate having signed the message, since no one but the
signer possesses his private key. By analogy with the paper world,
where one might sign a letter and seal it in an envelope, one can
sign an electronic message using one's private key, and then seal
the result by encrypting it with the recipient's public key. The
recipient can perform the inverse operations of opening the letter
and verifying the signature to electronic mail are quite widespread
today already (see [GoB])

A digital signature scheme within the public key framework, is
defined as a triple of algorithms $(A, \sigma, V)$ such that

\begin{itemize}
    \item Key generation algorithm $A$ is a probabilistic, polynomial-time algorithm which on input a security parameter $1^k$, produces pairs $(P, S)$ where $P$ is called a public key and $S$ a secret key. (We use the notation $(P, S) \in G(1^k)$ to indicate that the pair $(P, S)$ is produced by the algorithm $A$.)

  \item Signing algorithm $\sigma$ is a probabilistic polynomial time algorithm which is given a security parameter $1^k$, a secret key $S$ in range $A(1^k)$, and a message $m \in \{0, 1\}^k$ and produces as output string $s$ which we call the signature of $m$. (We use notation $s \in \sigma(1^k, S, m)$ if the signing algorithm is probabilistic and $s = \sigma(1^k, S, m)$ otherwise. As a shorthand when the context is clear, the secret key may be omitted and we will write $s \in \sigma(S,m)$ to mean that s is the signature of message $m$.)

  \item Verification algorithm $V$ is a probabilistic polynomial time algorithm which given a public key $P$, a digital signature $s$, and a message $m$, returns $1$ (i.e "true") or $0$ (i.e "false") to indicate whether or not the signature is valid. We require that $V (P, s, m) = 1$ if $s \in \sigma(m)$ and $0$ otherwise. (We may omit the public key and abbreviate $V(P, s, m)$ as $V(s, m)$ to indicate verifying signature $s$ of message $m$ when the context is clear.)

  \item The final characteristic of a digital signature system is its security against a probabilistic polynomial time forger. We delay this definition until later.
\end{itemize}

We present a digital signature procedure based on nonabelian
groups developed by Ko, Lee et al(see
[KCCL]. Here is the scheme:

Let $G$ be a non-abelian group in which the search conjugacy problem is infeasible and the decision conjugacy problem is solvable.  Let $h: \left\{0,1\right\}^{*} \rightarrow G$ be a hash function.\\

\textbf{\textit{Key Generation:}} Alice wants to sign and send a
message, $m$, to Bob. Alice begins by choosing
 two conjugate elements $u,v\in G$ with conjugator $a$. The conjugate pair $(u,v)$ is public information while the conjugator $a$ is Alice's secret key. \\

\textbf{\textit{Signature Generation:}} Alice chooses arbitrary $b\in G$, and computes $\alpha=u^{b}$ and $y=h(m\alpha)$.  Then a signature $\sigma$ on the message $m$ is the triple $(\alpha, \beta, \gamma)$ where $\beta=y^{b}$ and $\gamma=y^{a^{-1}b}$. She sends this to Bob for verification and acceptance.\\

\textbf{\textit{Verification:}} Upon receiving the signature, Bob
checks whether or not the following hold:
\begin{enumerate}
\item $\exists c_{1}\in G$ such that $u=\alpha^{c_{1}}$.
\item $\exists c_{2}, c_{3} \in G$ such that $\gamma= \beta^{c_{2}}$ and $y=\gamma^{c_{3}}$.
\item $\exists c_{4}\in G$ such that $uy=(\alpha\beta)^{c_{4}}$.
\item $\exists c_{5}\in G$ such that $vy=(\alpha\gamma)^{c_{5}}$.
\end{enumerate}
Bob accepts the signature if and only if 1-4 hold.

The security of this scheme lies in the assumption that given a pair of conjugate elements $u,v \in G$ finding elements $\alpha, \beta, \gamma$ such that 1-4 above hold is infeasible. If the conjugator $a$ could be found then, then $(\alpha, \beta, \gamma)=(u^{b},y^{b}, y^{a^{-1}b})$ satisfy properties 1-4 for any $b\in G$.  Hence, the conjugacy search problem need be infeasible.\\

We mention that there is digital signature scheme proposed by Anjaneyulu et. al [AVRR] that uses a noncommutative platform but outside of group theory.  In this proposal the basic cryptographic platform is a commutative division semiring and uses what is teremed the polynomial syemmtrical decompisiton problem as the one way function.  The reader can refer to the paper [AVVR] for details.

Closely related to digital signatures is the problem of \textbf{secure password verification}. With the increased use of bank cards and internet credit card transactions there is at present more than ever a need for secure password identification. For many online purchases this is being carried out by a {\bf challenge response system} (see [W]) accompanying the password. In the simplest systems this takes the form of secondary password questions such as the user's mother's maiden name or place of birth.  There are inherent difficulties with these types of challenge response systems. First of all there is the trivial problem of the users remembering their responses.  More critical is the problem that this type of information for many people is readily available and easily found or guessed by would-be attackers or eavesdroppers. Challenge response systems are also subject to middleman attacks and replay attacks (see [CRW].  There have been several attempts to alleviate these problems, including zero-knowledge password proofs and challenged responses somewhat based on RSA as well as timed out responses (see CRAM-MD5, Password Authenticated Key Agreement, [CRRSA] and [W]).

In [BBFT] an alternative method for challenge response password verification using combinatorial group theory was developed which is provably secure against cipher-text only, man in the middle and replay attacks. In particular this method depends upon the difficulty of solving the word problem within a given finitely presented group without knowing the presentation and the difficulty of solving systems of equations within free groups. This latter problem has been proved to be NP-hard. The method uses the \textbf {group randomizer system} which is a computer program that is a subset of MAGNUS a much larger computer algebra system designed to handle algorithm problems in combinatorial group theory, MAGNUS was developed at CAISS, the Center for Algorithms and Interactive Scientific Software, a research laboratory housed at City College of the City University of New York and under the direction of the first author. The group randomizer system can be placed on a simple hand held computer device presently under development at CAISS. The system can also be used from computer to computer (see [BBFT]).

These group theoretic techniques have several major advantages over other challenge response systems. We will call the password presenter the \textbf {prover} and the presentee the \textbf {verifier}.  The methods we present can be used for \textbf{two-way authentication}. That is the same method can both authenticate the prover to the verifier and authenticate the verifer to the prover.  To each prover in conjunction with a standard password there will be assigned a finitely presented group with a solvable word problem. This is the {\bf challenge group}. This will be done randomly by the group randomizer system and will be held in secret by the prover and the verifier.  Cryptographically we assume the adversary can steal the encrypted form of the group theoretic responses.  Probabilistically this does not present a problem. Each challenge response set of questions forms a virtual one time key pad.  Therefore the adversary must steal three things - the original password, the challenge group and the group randomizer. Hence there is almost total security in the challenge response system. Further there is an infinite supply of finitely presented groups to use as challenge groups and an infinite supply of challenge response questions that never have to be duplicated.

The theoretical security of the system is provided by several results in asymptotic group theory.  In particular a result of Lysenok [L] implies that stealing the challenge group is NP-hard while a result of Jitsukawa [J] says that the asymptotic density of using homomorphisms to attack the group randomizer protocol is zero (see sections 9 and 10)

In brief outline the system works as follows. We assume that each prover has a group randomizer system.  At the most basic level the group randomizer system has the ability to do the following things:

$\hphantom{xx}$ (1) Recognize a finite presentation of a finitely presented group with a solvable word problem and manipulate arbitrary words in the alphabet of generators according to the rewriting rules of the presentation.  In particular if the group is automatic the group randomizer can rewrite an arbitrary word in the generators in terms of its group normal form.

$\hphantom{xx}$ (2) Given a finite presentation of a group with a solvable word problem recognize whether two free group words have the same value in the given group when considered in terms of the given generators of the group

$\hphantom{xx}$ (3) Randomly generate free group words on an alphabet of any finite size

$\hphantom{xx}$ (4) Recognize and store sets of free group words $W_1,...,W_k$ on an alphabet $x_1,...,x_n$ and rewrite words $W(W_1,...,W_k)$ as the corresponding word in $x_1,...,x_n$.

$\hphantom{xx}$ (5) Given a free group of finite rank on $x_1,...,x_n$ and a set of words $W_1,...,W_k$ on an alphabet $x_1,..,x_n$ solve the membership problem in $F$ relative to $H = <W_1,...,W_k>$, the subgroup of $F$ generated by $W_1,...,W_k$.

$\hphantom{xx}$ (6) Given a stored finitely presented group or a stored set of free group words the randomizer can accept a random free group word and rewrite it as a normal form in the finitely presented group in the former case or as a word in the ambient free group in the latter case.

 Each prover further has a standard password.  Suppose that $F$ is a free group on $\{x_1,...,x_n\}$. The prover's password is linked to a finitely generated subgroup of a free group given as words in the generators - that is the prover's password is linked to $W_1,...,W_k$ where each $W_i$ is a word in $x_1,...,x_n$. The group $G = <W_1,...,W_k>$ is called the {\bf challenge group}. In general $k \ne n$.  The prover doesn't need to know the generators. The randomizer can randomly choose words from this subgroup  and then freely reduce them.  The verifier has the challenge group or subgroup also stored in its randomizer. From the viewpoint of cryptology this is a symmetric key protocol with both prover and verifier having a common shared secret, $(P,G)$, where $P$ is a standard password and $G$ is the challenge group. The shared secret is set at initialization of the protocol by some direct communication. This is the most common model for password security.

The prover submits his or her standard password to the verifier. This activates the verifier's randomizer to the prover's set of words.  The verifier now submits a random free group word on $y_1,...,y_k$ to the prover's randomizer say $W(y_1,...,y_k)$. The prover's randomizer treats this as $W(W_1,...,W_k)$ and then reduces it in terms of the free group generators $x_1,...,x_n$ and rewrites it as $W^\star(x_1,...,x_n)$.  The verifier checks that this is correct - that is $W(W_1,...,W_k) = W^\star(x_1,...,x_n)$ on the free group on $x_1,...,x_n$.  If it is the verifier continues and does this three (or some other finite number) of times. There is one proviso. The verifier submits a word to the prover only once so that a submitted word can never be reused.  The prover's randomizer will recognize if it has (this is a verification to the prover of the verifier).

To verify that the verifier is legitimate the process is repeated from the prover's randomizer to the verifier.

In [BBFT] several variations of this basic outline using more general finitely presented groups were presented. Using some results from asymptotic group theory it was shown that this method is provably secure.

\section{Braid Group Cryptography and Platform Groups}

As platform groups for their respective protocols, both Ko-Lee and Anshel-Anshel-Goldfeld suggested the braid groups $B_n$ (see [Bi]).  The groups in this class of groups possess the desired properties for the key exchange and key transport protocols; they have nice presentations with solvable word problems and conjugacy problems; the solution to the conjugacy and conjugator search problem is "hard"; there are several possibilities for normal forms for element and they have many choices for large commuting subgroups.  Initially the braid groups were considered so ideal as platforms that many other cryptographic applications were framed within the braid group setting.  These included {\bf authentication} and {\bf digital signatures}.  There was so much enthusiasm about using these groups that the whole area of study was named {\bf braid group cryptography}. A comprehensive and well-written article by Dehornoy [D] provides a detailed overview of the subject and we refer the reader to that for technical details.

After the initial successes with braid group cryptographic schemes there were some surprisingly effective attacks.  There were essentially three types of attacks; an attack using solutions to the conjugacy and conjugator search problems, an attack using heuristic probability within $B_n$ and an attack based on the fact that there are faithful linear representations of each $B_n$ (see [D]).  What is most surprising is that the Anshel-Anshel-Goldfeld method was susceptible to a length based attack. In the Anshel-Anshel-Goldfeld method the {\bf parameters} are the specific braid group $B_n$ and the rank of the secret subgroups for Bob and Alice. A length based attack essentially broke the method for the initial parameters suggested by Anshel, Anshel and Goldfeld in [AAG]. The parameters were then made larger and attacks by this method were less successful.  However this led to research on why these attacks on the conjugator search problem within $B_n$ were successful,  What was discovered was that {\bf generically} (see sections 9 and 10) a random subgroup of $B_n$ is a free group and hence length based attacks are essentially attacks on free group cryptography and therefore successful.  What this indicated was that although randomness is important in cryptography in using the braid groups as platforms subgroups cannot be chosen purely randomly.

Braid groups arise in several different areas of mathematics and have several equivalent formulations. What we do in the remainder of this section is describe the braid groups. A complete topological and algebraic description can be found in the book of Joan Birman [Bi].

A {\bf braid} on $n$ strings is obtained by starting with $n$ parallel strings and intertwining them. We number the strings at each vertical position and keep track of where each individual string begins and ends. We say that two braids are equivalent if it is possible to move the strings of one of the braids in space without moving the endpoints or moving through a string and obtain the other braid.   A braid with no crossings is called a {\bf trivial braid}. We form a product of braids in the following manner. If $u$ is the first braid and $v$ is the second braid then $uv$ is the braid formed by placing the starting points for the strings in $v$ at the endpoints of the strings in $u$. The inverse of a braid is the mirror image in the horizontal plane. It is clear that if we form the product of a braid and its mirror image we get a braid equivalent to the trivial braid. With these definitions, the set of all equivalence classes of braids on $n$ strings forms a group $B_n$.  We let $\sigma_i$ denote the braid that has a single crossing from string $i$ over string $i+1$. Since a general braid is just a series of crossings it follows that $B_n$ is generated by the set $\sigma_i; i = 1,...,n-1$.

There is an equivalent algebraic formulation of the braid group $B_n$.  Let $F_n$ be a free group on the $n$ generators $x_1,...,x_n$ with $n > 2$. Let $\sigma_i,i = 1,...,n-1$ be the automorphism of $F_n$ given by
$$ \sigma_i:x_i \mapsto x_{i+1}, x_{i+1} \mapsto x_{i+1}^{-1}x_ix_{i+1}$$
$$ \sigma_i;x_j \mapsto x_j,j \ne i,i+1.$$
Then each $\sigma_i$ corresponds precisely to the basic crossings in $B_n$.  Therefore $B_n$ can be considered as the subgroup of $Aut(F_n)$ generated by the automorphisms $\sigma_i$,  Artin proved [A] (see also [MKS]) that a finite presentation for $B_n$ is given by
$$B_n = <\sigma_1,...,\sigma_{n-1}; [\sigma_i,\sigma_j] = 1 \text { if } |i-j|>1, x_{i+1}x_ix_{i+1} = x_{i}x_{i+1}x_i, i = 1,...,n-1>.$$
This is now called the Artin presentation.  The fact that $B_n$ is contained in $Aut(F_n)$ provides an elementary solution to the word problem in $B_n$ since one can determine easily if an automorphism of $F_n$ is trivial on all the generators. We note that although the braid groups $B_n$ are linear ( the Lawrence-Krammer representation is faithful (see[D]) it is known that $Aut(F_n)$ is not linear (see [F]).

There are several possibilities for normal forms for elements of $B_n$.  The two most commonly used are the {\bf Garside normal form} and the {\bf Dehornoy normal form}.  These are described in [D] and [MSU].

From the commuting relations in the Artin presentation it is clear that each $B_n$ has the requisite collection of commuting subgroups.

The conjugacy problem for $B_n$ was originally solved by Garside and it was assumed that it was hard in the complexity sense.  Recently there has been significant research on the complexity of the solution to the conjugacy problem (see [MSU] and [D]).

In general, platform groups for the non-commutative protocols that we have discussed require certain
properties.  Most are present in the braid groups.  The first is the existence of a normal form for elements in the group.
Normal forms provide an effective method of disguising
elements.  Without this, one can determine a secret
key simply by inspection of group elements. The existence of a
normal form in a group implies that the group has solvable word
problem, which is essential for these protocols. For purposes of
practicality, the group also needs an efficiently computable normal
form, which ensures an efficiently solvable word problem.

In addition to the platform group having normal form, ideally, it
would also exhibit exponential growth.  That is, the growth function
for $G$, $\gamma: \mathbb{N} \rightarrow \mathbb{R}$ defined by
$\gamma(n)=\#\left\{w\in G : l(w)\leq n\right\}$, has an exponential
growth rate.  Exponential growth is a necessity since this ensures
that the group will provide a large key space, making a brute force
search for the secret key an infeasible algorithm.

The other property which is necessary in most of the proposed
cryptosystems is the conjugator search problem for the platform
group ideally should have exponential time complexity.

In addition to these, for the Ko-Lee type protocols, we need large commuting subgroups wihtin the platform group.

Currently, there are many potential platform groups that have been suggested.
 The following are some of the proposed platform groups:
\begin{itemize}
    \item Braid groups(Ko-Lee, Anshel-Anshel-Goldfeld)
    \item Thompson Groups (Shpilrain-Ushakov) [SU]
    \item Polycyclic Groups (Eick-Kahrobaei) [EK]
    \item Linear Groups (Baumslag-Fine-Xu) [BFX 1,2]
    \item Free metabelian Groups (Shpilrain-Zapata) [SZ]
    \item Grigorchuk Groups (Petrides) [P]
    \item Groups of Matrices (Grigoriev-Ponomarenko) [GP]
    \item Surface Braid Groups (Camps) [C]
\end{itemize}

Many of these are discussed in [MSU].

\section{Cryptography With Polycyclic Groups}

In this section we briefly discuss polycyclic group cryptography which has not been extensively studied but has many of the essential features for ideal platform groups. A cryptosystem using polycyclic groups was developed in [EK].

A group $G$ is called polycyclic if it has a series  $G=G_{n+1}\geq G_{n}\geq \cdots \geq G_{2}\geq G_{1}={1}$ in which each $G_{i}$ is a
normal subgroup of $G_{i+1}$ and $G_{i+1}/G_{i}$ is cyclic for
$i=1,2,\cdots, n$. A series of this type is called a polycyclic
series. Polycyclic groups are a natural non-commutative generalization of
cyclic groups. The book of Holt et,al. [HEO] is a good reference for information about polycyclic
groups.

Every polycyclic group $G$ has a finite presentation of the form:
\begin{eqnarray*}
\left\langle a_{1}, \cdots, a_{n}|a_{i}^{a_{j}}=w_{ij},
a_{i}^{a_{j}^{-1}}=v_{ij},  a_{k}^{r_{k}}=u_{kk}\right\rangle
\end{eqnarray*}

for $1\leq j< i \leq n$ where $r_{i}\in \mathbb{N}\cup \infty$,
$r_{i}< \infty$ if $i\in I\subseteq\left\{1, 2,\cdots, n\right\}$
and $w_{ij}, v_{ij}, u_{jj}$ are words in the generators $a_{j+1},
\cdots, a_{n}$.  If $r_{i}=[G_{i+1}:G_{i}]$ for each $i\in \left\{1,
2,\cdots, n\right\}$ then this presentation is called a consistent
polycyclic presentation.  Every element in the group defined by this
consistent polycyclic presentation may be written uniquely in the
form $a_{1}^{e_{1}} \cdots a_{n}^{e_{n}}$ with $e_{i}\in \mathbb{Z}$
and $0\leq e_{i}<r_{i}$ if $i\in I$.  This unique representation of
each element $g\in G$ is called the normal form of $G$. It is known
that every polycyclic group exhibits a consistent polycyclic
presentation.  Hence, every polycyclic group has a normal form. This
is used as a basis for computations with polycyclic groups.

The word problem can be solved effectively using the collection
algorithm in a group $G$ given by a consistent polycyclic
presentation. The collection algorithm computes the unique normal
form for an element, $g$, in the group given by a word in the
generators.  This is done by repeatedly applying the power and
conjugacy relations given in the presentation to subwords of  $g$,
transforming $g$ to an equivalent word.  The nature of the relations
ensures that this process must terminate, producing the unique
normal form for $g$.  The collection algorithm is known to be a
practical and effective method for solving the word problem in
consistent polycyclic presentations.

Every polycyclic group can be embedded in  $GL(n,\mathbb{Z})$, which
reveals important properties about polycyclic groups. Since matrix
multiplication is solvable in polynomial time, group multiplication
in polycyclic groups is efficient. As polycyclic groups have a
normal form, efficiently solvable group multiplication implies that
the word problem is also efficiently solvable. It has been proven
that the search conjugacy problem in any subgroup of a general
linear group is solvable. Because every polycyclic group can be
embedded as a subgroup of $GL(n,\mathbb{Z})$, the search conjugacy
problem in polycyclic groups is solvable. The complexity of the
search conjugacy problem in polycyclic groups is unknown, but widely
conjectured to be exponential time.

Recall that the Anshel-Anshel-Goldfeld key exchange can be broken if
the simultaneous search conjugacy problem is solvable.  In
polycyclic groups, the simultaneous search conjugacy problem reduces
to the search conjugacy problem.

\section{Generic Complexity and Asymptotic Density}

In cryptanalysis involving group theoretic decision problems what is important is not just the solution to the problem but the computational complexity, polynomial or exponential for example,  of the algorithm to solve the problem. The problem may be hard  or even undecidable on some inputs but actually easy on most inputs.  This is in reality a problem in many braid group schemes.  The conjugator search problem is hard on some inputs but actually easy for many chosen subgroups.  A problem may be very hard on some inputs.  This is called {\bf worst case complexity}. More important for cryptanalysis is {\bf average case complexity}, that is the complexity on average over all inputs. {\bf Generic complexity} refers to the complexity of the solution to a particular algorithm over most inputs.  Generic and average case complexity and their uses in cryptanalysis are discussed in detail in [MSU].  There they show that generic complexity is a more useful tool in most cryptographic applications.  There they show that in many cases if an algorithm is easy on average it is also easy generically.  They also show that the opposite is not true and provide examples of algorithms that are exponential on average but polynomial time generically.

The problem with some braid group cryptographic algorithms is that random subgroups of $B_n$ are generically free.  We now describe what this means.

Asymptotic density is a general method to compute densities and/or probabilities on infinite discrete sets where each individual outcome is tacitly assumed to be equally likely. The method can also be used where some probability distribution is assumed on the elements.  It has been effectively applied to determining densities within infinite discrete finitely generated groups where random elements are considered as being generated from random walks on the Cayley graph of the group.  The paper by Borovik, Myasnikov and Shpilrain [BMS] provides a good general description of this method in group theory. Let $\mathcal P$ be a group property and let $G$ be a finitely generated group. We want to determine the measure of the set of elements which satisfy $\mathcal P$.  For each positive integer $n$ let $B_n$ denote the $n$-ball in $G$. Let $|B_n|$ denote the actual size of $B_n$ (which is an integer since $G$ is finitely generated) or the measure of $|B_n|$ if a distribution has been placed on the elements of $G$. Let $S$ be the set of elements in $G$ satisfying $\mathcal P$.  The asymptotic density of $S$ is then
$$ \lim_{n \to \infty} \frac{|S \cap B_n|}{|B_n|}$$
provided this limit exists.  We say that the property $\mathcal P$ is {\bf generic} in $G$ if the asymptotic density of the set $S$ of elements satisfying $\mathcal P$ is one, ${\cal P}$ is called an \textbf{asymptotic visible} property, if the corresponding asymptotic density is strictly between $0$ and $1$. If the asymptotic density is $0$, then ${\cal P}$ is called \textbf{negligible}.

This concept can be easily extended to properties of finitely generated subgroups, We consider the asymptotic density of finite sets of elements that generate subgroups that have a considered property.  For example to say that a group has the generic free group property we mean that
$$ \lim_{m,n \to \infty} \frac{|S_{m} \cap B_{m,n}|}{|B_{m,n}|} = 1$$
where $S_m$ is the collection of finite sets of elements of size $m$ that generate a free subgroup and $B_{m,n}$ is the collection of $m$ element subsets within the $n$-ball.

If $\mathcal P$ is a group property and $G$ is a group then we say that subgroups of $G$ are {\it generically }$\mathcal P$ if a generic randomly chosen subgroup $H$ of $G$ has property $\mathcal P$. Equivalently this means that the asymptotic density  of subgroups $H$ of $G$ that have property $\mathcal P$ is one.

\section{The Generic Free Group Property}

In general we say that a group $G$ has the {\bf generic free group property} if a finitely generated subgroup is generically a free group.
A result of Epstein [E] shows that the group $GL(n,\Bbb R)$ satisfies the generic free group property.  Further $G$ has the {\bf strong generic free group property} if given randomly chosen elements $g_1,...,g_n$ in $G$ then generically they are a free basis for the free subgroup they generate.  Jitsukawa [J] showed that finitely generated nonabelian free groups have the strong generic free group property while Gilman, Myasnikov and Osin [GMO] showed that torsion-free hyperbolic groups also have the generic free group property.
Myasnikov and Ushakov [MU] showed that pure braid groups $P_n$ with $n \ge 3$ also have the strong generic free group property. A recent result of Carstensen, Fine and Rosenberger [CFR] shows that all Fuchsian groups of finite co-volume and all braid groups $B_n$with $n \ge 3$ have the strong generic free group property,

This result of Myasnikov and Ushakov on the pure braid groups has applications to the cryptanalysis of both the Ko-Lee cryptosystem and the Anshel-Anshel-Goldfeld cryptosystem (see [SU] and [MU]). Both cryptosystems were usually suceptible to length based attacks if the parameters chosen in the braid groups $B_n$ were small.  The reason for this is that random choices of subgroups within the braid groups are actually free groups.  This does not disqualify the braid groups as platforms but rather says that subgroups cannot be chosen entirely randomly.

Extremely useful in proving that a group has the generic or strong generic free group property is the following.

\begin{theorem} Let $G$ be a group and $N$ a normal subgroup.  If the quotient $G/N$ satisfies the strong generic free group property then $G$ also satisfies the strong generic free group property.
\end{theorem}

In [FMR] it was shown that many group amalgams - free products, free products with amalgamtion and HNN groups, satisfy the strong generic free group property.  In particular the most general result is the following.

\begin{theorem} Let $A$ and $B$ be arbitrary finitely generated infinite groups and let $G = A \star B$ be their free product.  Let $\{x_1,...,x_n\}$ be $n$ randomly chosen elements from $G$. Then generically these elements are a free basis for the subgroup they generate, that is $G$ satisfies the strong generic free group property.
\end{theorem}

This can be extended to more general amalgams in many ways (see [FMR])

\begin{theorem} Let $A$ and $B$ be arbitrary finitely generated infinite groups and let $G = A \underset H \star B$ be their amalgamated free product with amalgamated subgroup $H$. Let $H_1$ and $H_2$ be the copy of $H$ in $A$ and $B$ respectively. Suppose that $A/N(H_1)$ is infinite and $B/N(H_2)$ is infinite where $N(H_i)$ is the normal closure of $H_i$ in the respective factors.  Then $G$ satisfies the strong generic subgroup property.
\end{theorem}

Recall that a {\bf cyclically pinched one-relator group} is a amalgamated free product of the form
$$ G = F_1 \underset {\{U = V\}} {\star} F_2$$
where $F_1,F_2$ are finitely generated free groups and $U,V$ are nontrivial words in the respective free groups. If $U$ is not a power of a primitive element in $F_1$ and $V$ is not a power of a primitive element in $F_2$ then the quotient of $F_1$ and $F_2$ by the normal closure of $U$ and $V$ respectively is a nontrivial, infinite one-relator group.

\begin{corollary} Let $G$ be a cyclically pinched one-relator group as above. Assume that $U$ and $V$ are not a  power of a primitive element in $F_1$ and $F_2$ respectively.  Then $G$ satisfies the strong generic subgroup property.
\end{corollary}

In particular any orientable surface group of genus $g \ge 2$ falls into the class of cyclically pinched one-relator groups.

\begin{corollary} Any orientable surface group of genus $g \ge 2$ and any nonorientable surface group of genus $g \ge 4$ satisfies the strong generic subgroup property.
\end{corollary}

The situation with HNN groups becomes even more complicated but some things can be proved as consequences of the amalgam result above. Notice first however that any HNN group with free part of rank $\ge 2$ must have a free quotient of rank $\ge 2$ and hence satisfy the strong generic subgroup property. Therefore only the case where the free part has rank 1 must be considered.

\begin{theorem} Let $G$ be an HNN extension of the group $B$ with a presentation
$$ G = < t,B; rel(B), t^{-1}Ut = V>$$
with $U,V$ nontrivial isomorphic subgroups of $B$.  Let $N_B(<U,V>)$ be the normal closure of the subgroup $<U,V>$ in $B$. Then if $B/N_B(<U,V>)$ is infinite, $G$ satisfies the strong generic subgroup property.
\end{theorem}

Extensions of centralizers play a large role in the study of the elementary theory of free groups.  Recall that if $B$ is a group and $U \in B$ then a {\bf rank one extension of centralizers} of $B$ is a group with a presentation
$$G = <t,B; rel(B), t^{-1}UT = U>.$$

\begin{theorem} Let $G$ be a rank one extension of centralizers of the group $B$.  Suppose $G$ has a presentation
$$ G = < t,B; rel(B), t^{-1}Ut = U>$$
wihere $U$  is a nontrivial element of $B$. If $B/N_B(U)$ is infinite, where $N_B(U)$ is the normal closure of $U$ in $B$, then $G$ satisfies the strong generic subgroup property .
\end{theorem}

In the situation where the factors are finite we must be careful even for free products.  The infinite dihedral group $\Bbb Z_2 \star \Bbb Z_2$ is solvable so cannot satisfy the strong generic free group property. However if at least one factor has order greater than 2, an analysis based on Kurosh bases yields the weaker generic free group property.

\begin{theorem} Let $G = A \star B$ be a nontrivial free product.  If at least one factor has order greater than 2 then $G$ satisfies the generic free group property.
\end{theorem}

In [CFR] it was shown that a finitley generated group satisfies the strong generic free group property if and only if subgroups of fintie index do also. We call a group property $\cal{P}$ {\bf suitable} for a finitely generated group $G$ if it is preserved under isomorphisms and its asymptotic density is independent of finite generating systems.  From a result in [MSU] the strong generic free group property is suitable in any group that has a nonabelian free quotient.

The main result in [CFR] is the following called the {\bf inheritance theorem}.

\begin{theorem} Let $G$ be a finitely generated group and $H<G$ a subgroup of finite index $[G:H]=n<\infty$. Let ${\cal P}$ be the strong generic free group property. Then:
\begin{enumerate}
\item If ${\cal P}$ is a suitable and generic property in $H$ then it is also suitable and generic in $G$.
\item If ${\cal P}$ is a suitable and generic property in $G$ then it is also suitable and generic in $H$.

\end{enumerate}
\end{theorem}

An interesting consequence of this is that all braid groups have the strong generic free group property.

\section{Open Problems}

We now give a nonexhaustive list of problems related to the rest of this article.

\textbf{General:}

\begin{enumerate}
  \item What is the  most appropriate platform group for
  non-commutative cryptography?
  \item Should the group the be finite or infinite?
  \item How can we show a group is provably secure for the new
  non-commutative schemes such as public key exchanges, digital
  signatures and authentication?
  \item Can we design more public keys based on other search and
    decision problems in combinatorial group theory?
    \item Can we analyze the security of this protocols?
    \item what should be the measure of the security? (practicality,
    complexity, average case complexity, generic complexity?)
    \item So far there are three known non-commutative digital
    signatures have been designed, can we design more
    non-commutative digital signatures?
    \item What about the authentication schemes?
    \item What is the appropriate choice of commuting subgroups in
    polycyclic groups which makes the described schemes secure?
\end{enumerate}

\textbf{Complexity Analysis and Security:}

\begin{enumerate}
    \item What is the complexity of the search conjugacy problem in
    polycyclic groups?
    \item What is the complexity of the decision conjugacy problem,
    for example in Eick-Ostheimer algorithm [EO]?
    \item What is the complexity of the collection algorithm to find
    the normal form of elements in polycyclic groups?
    \item What is the complexity of the twisted search conjugacy
    problem in polycyclic groups?
\item What is the complexity of the Reidemeister-Schreier rewriting algorithm in free groups?
\item What is the generic case complexity of the search
    conjugacy problem in polycyclic groups?
\end{enumerate}

\textbf{Quantum Algorithms and Quantum Complexity:}

Another problem to think in this direction is quantum computational
approaches to these cryptosystems. Quantum
algorithms for finite solvable groups (which are polycyclic) has
been studied, particularly by J. Watrous (2001) [Wa]. He
found a quantum algorithm to compute the order of a finite solvable
group in polynomial time. Algorithm works in the setting of
black-box groups none of them have polynomial-time classical
algorithms. Can we design quantum algorithms for solving other
decision problems in polycyclic groups (both for finite and infinite
ones); especially the ones we use in cryptography.

\begin{enumerate}
    \item Is there any quantum algorithm for solving the search
    conjugacy problem for polycyclic group that reduces the
    complexity of the algorithm?
\end{enumerate}

\textbf{Implementation:}
\begin{enumerate}
    \item How can we implement the proposed cryptosystems?
    \item Can the computer algebra system GAP,
    could we use this for practical and secure cryptography?
\end{enumerate}

\section{Acknowledgment}
Benjamin Fine has been supported by a Gambrinus Grant. Delaram Kahrobaei has been supported by a grant from the research foundation of CUNY (PSC-CUNY) and the Faculty Fellowship Publication Program Award from
City University of New York.

\section{References}

\noindent \lbrack AVRR] G.S.G.N.Anjaneyulu and P.Vasudeva Reddy and U.M.Reddy, {\it Secured
Digital Signature Scheme using Polynomials over Non-Commutative
Division Semirings }, \textbf{IJCSNS - International Journal of Computer
 Science and Network Security}, 8, (2008), 278--284.

\noindent \lbrack AAG] I.Anshel,M.Anshel,D.Goldfeld, {\it An Algebraic Method for Public Key
Cryptography}, \textbf{ Math.Res.Lett}, 6, 1999,  287-291  Springer Verlag

\noindent \lbrack AK] M.Anshel and D.Kahrobaei, {\it A Noncommutative Analog of the Cramer-Shoup Public Key Exchange Protocol},
\textbf{ Groups, Complexity, Cryptology,}1, 2009, 217-225

\noindent \lbrack Ar] G.Arzhanseva, {\it Generic Properties of Finitely Presented Groups and Howson's Theorem}, \textbf{ Comm. Alg,}  26, 1998, 3783-3792

\noindent \lbrack AO] G.Arzhanseva and A. Olshanskii, {\it Genericity of the Class of Groups in Which Subgroups with a Lesser Number of Generators are Free } \textbf{Mat. Zametki}, 59, 1996, 489-496

\noindent \lbrack BBDR] M. Batty, S. Braunstein, A. Duncan, S. Rees, {\it Quantum algorithms in
group theory}, \textbf{Cont. Math. } 349, 2003, 1--62.

\noindent \lbrack GB 1] G.Baumslag, \textbf{ Topics in Combinatorial Group Theory},Birkhauser  1993

\noindent \lbrack BBFR] G. Baumslag, Y.Brjukhov, B.Fine and G.Rosenberger {\it Some Cryptoprimitives for Noncommutative Algebraic Cryptography }  \textbf{Aspects of Infinite Groups}, World Scientific Press, 2009, 26-44

\noindent \lbrack BCFRX] G. Baumslag, Y.Brjukhov, B.Fine and D.Troeger, {\it Challenge Response Password Security Using Combinatorial group Theory } to appear \textbf{J.Groups, Complexity and Cryptology}

\noindent \lbrack BFX 1] G. Baumslag, B.Fine, and X.Xu, {\it Cryptosystems Using Linear Groups } \textbf{ Appl. Alg. in Engineering, Communication and Computing }  17, 2006,  205-217

\noindent \lbrack BFX 2] G. Baumslag, B.Fine,  and X.Xu,  {\it A Proposed Public Key Cryptosystem Using the Modular Group} \textbf{  Cont.Math. } 421, 2007, 35-44

\noindent \lbrack BCFRX] G. Baumslag,T.Camps, B.Fine,G.Rosenberger  and X.Xu, {\it Designing Key Transport Protocols Using Combinatorial Group Theory }, \textbf{ Cont. Math. } 418, 2006, 35-43

\noindent \lbrack Bi] J.Birman, \textbf{Braids, Links and Mapping Class Groups } Annals of Math Studies Vol. 82, Princeton University Press, 1975

\noindent \lbrack BMS] A.Borovik, A.G. Myasnikov and V.Shpilrain, {\it Measuring Sets in Infinite Groups } \textbf{Cont.Math. } 298, 2002, 21-42

\noindent \lbrack B] J.A. Buchmann, \textbf{ Introduction to Cryptography }, Springer 2004

\noindent \lbrack Ca] P.J. Cameron, {\it Aspects of infinite permutation groups}, in
  \textbf{Groups St Andrews 2005, Vol.1} (C.M. Campbell et al., eds.),
  London Math. Soc. Lecture Note Ser. \textbf{399} (CUP, Cambridge
  2007), 1--35

\noindent \lbrack C] T.Camps, {\it Surface Braid Groups as Platform Groups and Applications in Cryptography } PhD Dissertation Universitat Dortmund 2009

\noindent \lbrack CFR]  C. Carstensen, B.Fine and G.Rosenberger, {\it On Asymptotic Densities and Generic Properties in Finitely Generated Groups } \textbf{  preprint}

\noindent \lbrack CR] CRAM-MD5 Password Authenticated Key Agreement

\noindent \lbrack CRRSA] Challenge-Response System Based on RSA (http:www.cag.lcs.mit.edu/ rugina/ssh-procedures

\noindent \lbrack D] P.Dehornoy, {\it Braid-Based Cryptography } \textbf{ Cont. Math.}, 360, 2004,5--34

\noindent \lbrack EK] B.Eick, D.Kahrobaei, {\it Polycyclic groups: A new platform for cryptology? }
  \textbf{math.GR/0411077} (2004), 1--7.

\noindent \lbrack EO] B.Eick, G.Ostheimer, {\it On the orbit-stabilizer problem for integral matrix actions of polycyclic groups}, textbf{Math. Comp. (electronic) } 72, 2003, 1511--1529.

\noindent \lbrack E] D.B.A. Epstein, {\it Almost all Subgroups of Lie Group are Free } \textbf{ J. Alg.}, 19, 1971,  261-262

\noindent \lbrack FMR] B.Fine, A.Myasnikov and G.Roseberger, {\it Generic Subgroups of Amalgams } \textbf{J.Groups, Complexity and Cryptology}, 1, 2009, 51--61

\noindent \lbrack GMO] R.Gilman, A.G. Myasnikov and D.Osin, {\it Bounded Nielsen Property in Hyperbolic Groups } \textbf{ to appear }

\noindent \lbrack GP] D.Grigoriev and I. Ponomarenko, {\it Homomorphic Public-Key Cryptosystems Over
Groups and Rings } \textbf{ Quaderni di Matematica},  2005

\noindent \lbrack HGS] C.Hall, I. Goldberg, B. Schneier, {\it Reaction attaacks Against Several Public Key
Cryptosystems } \textbf{ Proceedings of Information and Communications Security ICICS 99 },
Springer-Verlag,  1999,  2-12

\noindent \lbrack HEO] D.Holt, B.Eick, E.O'Brien, \emph{Handbook of computational group theory}
  (Discrete Mathematics and its Applications (Boca Raton), Chapman and Hall/CRC 2005)

\noindent \lbrack J] T. Jitsukawa, {\it Malnormal Subgroups of Free Groups }, \textbf{Cont. Math.} 298, 2002, 83-96

\noindent \lbrack KKh] D.Kahrobaei and B.Khan, {\it A Non-Commutative Generalization of the ElGamal Key Exchange using Polycyclic Groups } \textbf{Proceeding of IEEE, GLOBECOM}, 2006, 1--5

\noindent \lbrack KCCL] K.Ko, D.Choi, M.Cho and J.Lee, {\it New signature scheme using conjugacy problem } \textbf{Cryptology ePrint Archive}, 168, 2002, 1--13.

\noindent \lbrack KoL] K.Ko, J.Lee, J.H. Cheon, J.W. Han, J.Kang, C.Park, {\it New Public-Key
Cryptosystem Using Braid Groups }, \textbf{Advances in Cryptology - CRYPTO 2000 Santa Barbara CA },
 Lecture Notes in Computer Science,  Springer  1880,  2000  166-183

\noindent \lbrack KM] I.Kapovich and A. Myasnikov, {\it Stallings Foldings and Subgroups of Free Groups } \textbf{J. Algebra } 248, 2003, 665-694.

\noindent \lbrack Ko] N.Koblitz, \textbf{Algebraic Methods of Cryptography},  Springer, 1998

\noindent \lbrack M]  W. Magnus, {\it Rational Representations of Fuchsian Groups and Non-Parabolic
Subgroups of the Modular Group },  \textbf{ Nachrichten der Akad Gottingen},  1973,  179-189

\noindent \lbrack MKS]  W. Magnus, A. Karass and D. Solitar \textbf{Combinatorial Group
Theory},  Wiley Interscience,New York,   1968

\noindent \lbrack MU] A.D.Myasnikov and A.Ushakov, {\it Length Based Attack and Braid Groups:Cryptanalysis of Anshel-Anshel-Goldfeld Key Exchange Protocol } \textbf{ Public Key Cryptography - PKC 2007 },  Lecture Notes in Computer Science, Springer , 2007

\noindent \lbrack MSU] A.G. Myasnikov, V.Shpilrain and A. Ushakov, \textbf{Group-Based Cryptography} Advanced Courses in Mathematics, CRM Barcelona, 2007

\noindent \lbrack MSU 1] A.G. Myasnikov, V.Shpilrain and A. Ushakov, {\it A Practical Attack on Some Braid Group Based Cryptographic Protocols }  \textbf{ CRYPTO 2005  Lecture Notes in Computer Science  3621}, 2005,  86-96

\noindent \lbrack P] G.Petrides, {\it Cryptanalysis of the Public Key System Based on the Grigorchuk Groups },  \textbf{Lecture Notes in Computer Science 2898}, springer Verlag,  2003, 234-244.

\noindent \lbrack SU] V.Shpilrain and A. Ushakov, {\it The Conjugacy Search Problem in Public Key Cryptography; Unnecessary and Insufficient } \textbf{ Applicable Algebra in Engineering, Communication and computing},  17, 2006  285-289

\noindent \lbrack SZ] V.Shpilrain and A. Zapata, {\it Using the Subgroup Memberhsip Problem in Public Key Cryptography } \textbf{Cont. Math. }  418, 2006, 169-179

\noindent \lbrack St] R. Steinwandt, {\it Loopholes in two public key cryptosystems using the modular
groups } \textbf{ preprint Univ. of Karlsruhle},  2000

\noindent \lbrack W] Challenge-response Authentication \textbf{Wikipedia, the free encyclopedia}, http://en.wikipedia.org/wiki/Challenge-response-authentication

\noindent \lbrack X] Xiaowei Xu, {\it Cryptography and Infinite Group Theory }, Ph.D. Thesis, CUNY, 2006

\noindent \lbrack Y] A.Yamamura, {\it Public Key cryptosystems using the modular group },  \textbf{ Lecture
Notes in Comput. Sci.} 1431, 1998,  203-216

\end{document}